\documentclass[journal]{IEEEtran}

\usepackage{amsmath,amsthm}
\usepackage{amssymb}
\usepackage{amscd}
\usepackage{textcomp}
\usepackage{hyperref}
\usepackage{fancyvrb}
\usepackage{wasysym}
\usepackage{tablefootnote}
\usepackage{adjustbox} 
\usepackage{array, makecell}
\usepackage{cellspace} %
\usepackage{listings}
\usepackage{cite}

\usepackage{color}
\definecolor{red}{rgb}{1,0,0}
\definecolor{blue}{rgb}{0,0,1}
\definecolor{green}{rgb}{0,0.5,0}
\definecolor{magenta}{rgb}{1,0,1}

\newsavebox{\ieeealgbox}

\newcommand{\be}{\begin{equation}}
\newcommand{\ee}{\end{equation}}

\newcommand{\US}{\mathtt{US}}
\newcommand{\GB}{\mathtt{GB}}

\newcommand{\rl}{c_{ab}}
\newcommand{\rc}{r_{ab}}
\newcommand{\rw}{r^{w}_{ab}}
\newcommand{\rs}{r^{\sigma}_{ab}}
\newcommand{\lre}{l_{ab}^{re}}
\newcommand{\ldc}{l_{ab}^{DC}}

\newcommand{\dist}{d^{re}_{ab}}

\newcommand{\cohesPost}{\rho_{ab}}

\definecolor{blue}{rgb}{0,0,1}

\definecolor{dgreen}{rgb}{0.0,0.5,0.0}

\definecolor{lightBlue}{rgb}{0.,0.5,0.5}

\begin{document}

\title{
Predicting Dynamic Stability from Static Features in Power Grid Models using Machine Learning
}

\author{
Maurizio Titz,
Franz Kaiser,
Johannes Kruse,
Dirk Witthaut
\thanks{The authors are with the 
Forschungszentrum J\"ulich, 
Institute for Energy and Climate Research - Systems Analysis and Technology Evaluation (IEK-STE), 52428 J\"ulich, Germany
and the 
Institute for Theoretical Physics, 
University of Cologne,
50937 K\"oln, Germany.}
}


\markboth{Journal of \LaTeX\ Class Files,~Vol.~13, No.~9, September~2014}%
{Shell \MakeLowercase{\textit{et al.}}: Bare Demo of IEEEtran.cls for Journals}

\maketitle

\begin{abstract}
A reliable supply with electric power is vital for our society. Transmission line failures are among the biggest threats for power grid stability as they may lead to a splitting of the grid into mutual asynchronous fragments. New conceptual methods are needed to assess system stability that complement existing simulation models.
In this article we propose a combination of network science metrics and machine learning models to predict the risk of desynchronisation events. Network science provides metrics for essential properties of transmission lines such as their redundancy or centrality. Machine learning models perform inherent feature selection and thus reveal key factors that determine network robustness and vulnerability.
As a case study, we train and test such models on simulated data from several synthetic test grids. We find that the integrated models are capable of predicting desynchronisation events after line failures with an average precision greater than $0.996$ when averaging over all data sets. Learning transfer between different data sets is generally possible, at a slight loss of prediction performance. Our results suggest that power grid desynchronisation is essentially governed by only a few network metrics that quantify the networks ability to reroute flow without creating exceedingly high static line loadings.
\end{abstract}


\IEEEpeerreviewmaketitle


\section{Introduction}\label{sec:into}

Modern society relies on a secure and stable supply with electric power, which makes a failure of the electric power system especially harmful \cite{lacommare2006cost,tab2011}. It is thus of utmost importance to make and keep it as fail proof as possible \cite{kroger2008critical}. The increased deployment of renewable power sources poses new challenges for power system stability due to their fluctuating nature \cite{anvari2016short,Staffell2018}. Furthermore, they are often built at places that offer favourable conditions for generation, far away from consumers, which increases grid loads \cite{pesch2014impacts,rodriguez2014transmission}. Similarly, energy intensive sectors, such as transport and heating, have to shift away from fossil fuels and towards electrification to reduce carbon emissions \cite{orths2019flexibility}, putting further strain on the power grid \cite{steinberg2017electrification,guminski2019system}. Grid stability in general is thus not only an important, but also a timely topic.

A central aspect of power system stability is synchronicity. All generators in a grid have to rotate in synchrony to guarantee a steady flow of electric power \cite{motter2013spontaneous,dorfler_synchronization_2013}. Stability is at risk if the grid is disturbed or damaged. In fact, most blackouts can be traced back to the outage of a single power system element such as a transmission line \cite{pourbeik2006anatomy}. In transmission grids, which are generally strongly meshed, this initial failure can then lead to a cascade due to overloads of other transmission lines. At some point, either directly or after some steps of the cascade, the grid becomes dynamically unstable and synchronicity is lost. This scenario occurred during the 2003 Italian power outage \cite{berizzi2004italian} and the 2006 Western European power outage \cite{UCTE07}. Notably, a desynchonization does not necessarily induce a large-scale blackout if the asynchronous fragments can be stabilised by control actions or load shedding. In fact, two such events were observed in the Continental European grid in 2021 \cite{entsoe2021a,entsoe2021b}.

To divert these catastrophes, grid operators have to be able to judge risks \emph{in time}. Warning systems that alert grid operators are in place, and risk regarding any single contingency can in principle be assessed via simulations (see \cite{balu1992line} for a review). However, the sheer number of potentially critical structural elements in power grid systems makes simulating all \emph{possible} contingencies \emph{in time} computationally impossible. Transmission grid operators thus have to in part rely on heuristics and experience. Modern machine learning methods may contribute to this assessment or the selection of relevant contingency cases to be simulated in detail \cite{wehenkel1997machine,alimi2020review}.

In this article we explore the capability of supervised machine learning to identify line failures that lead to a desynchronisation. We employ computationally cheap, static inputs that are readily available to the transmission grid operators such that models can be evaluated rapidly.
Furthermore, we focus on efficient models that are amenable to human interpretation and thus enable 
scientific insights  \cite{roscher2020explainable}.

\begin{figure*}[tb]
  \begin{minipage}[c]{0.67\textwidth}
    \includegraphics[width=\textwidth]{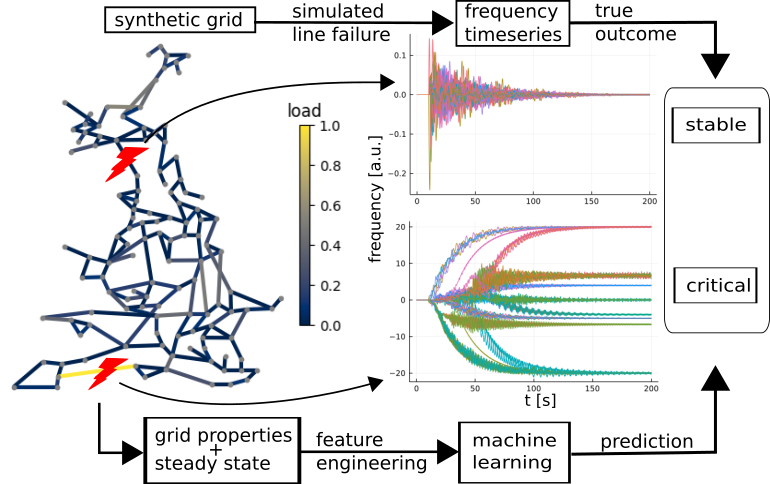}
  \end{minipage}\hfill
  \begin{minipage}[c]{0.3\textwidth}
    \caption{
       Summary of the machine learning model developed in this paper.
    We simulated the dynamics of a power grid after the failure of a single transmission line and classified the outcome as either stable or critical depending on whether nodal frequencies relaxed to zero. Input features for stability prediction are engineered from the properties of the grid and the pre-failure steady state using concepts from graph theory and network science. We then trained an explainable machine learning model to predict the stability outcome.
    } \label{fig:schema}
  \end{minipage}
\end{figure*}

A key method is the combination of machine learning with methods from graph theory and network science to develop features that are both understandable and have a high predictive power. Ideally, these feature will directly measure a relevant quantity of the respective line, such as its redundancy or centrality, which lend themselves to human interpretation. Machine learning models perform inherent feature selection and may thus reveal key factors that determine network robustness and vulnerability. Furthermore, they can incorporate different features to improve prediction. 

The article is organised as follows. In Sec.~\ref{sec:methods}, we will first give an overview over the model used for power grid dynamics and the resulting data sets. We then introduce the machine learning models, performance metrics and eXplainable Artificial Intelligence (XAI) methods used. After that, we introduce a multitude of different graph theory based features and assess their ability to quantify desynchronization risk in Sec.~\ref{sec:metrics}. We then go on to show that using machine learning produces improved predictions and quantify the contribution of different input features in Sec.~\ref{sec:ml-results}. Next, we demonstrate that the machine learning models are capable of learning transfer between different data sets. Finally, we interpret the predictions of one model by showing under which circumstances it tends to fail and how it outperforms simpler models.

\section{Methods}
\label{sec:methods}

Our approach is summarised in Figure \ref{fig:schema}. Using a coarse-grained model, we simulate the dynamics of a power grid after the failure of a single transmission line. We then classify the line according to the final post-failure state: The grid can either relax to a new synchronous steady state or lose synchronicity permanently. This procedure is repeated for all lines in a set of synthetic test grids, providing the raw data for the development of the machine learning models.

This work aims to predict the dynamic stability of a power system based solely on static features, i.e. features that can be derived from the pre-failure steady state, without making use of any simulation results. This way, our prediction method is computationally much cheaper than simulations, giving it a decisive advantage in an application case, were real time risk assessment with limited computational resources prohibits simulating all relevant contingencies. We make use of feature engineering based on concepts from graph theory and network science which will be described in detail in section \ref{sec:metrics}.

\subsection{Simulating the impact of line failures}

Models of varying complexity are used for power system stability analysis, depending on the scope of the analysis and the stability mechanism of interest \cite{machowski2020power,witthaut2022collective}. Since this work investigates solely the synchronisation behaviour on large spatial scales we focus on the voltage phase angles and frequencies, neglecting aspects of voltage stability or control. Furthermore, we focus on coarse spatial scales and thus consider aggregated models \cite{zhang1997sime,you2004slow,Fila08,chow2013power, MotterModels}. In this approach, each node of the network represents a small region or city, labelled by $i \in \{ 1,\ldots,N \}$. The dynamics of the local phase angle $\delta_i$ and frequency $\omega_i = \dot \delta_i$ is then determined by the aggregated swing equation
\begin{equation}
   J_i \ddot{\delta}_i + D_i \dot{\delta}_i = P_{i}^{(in)} - P_{i}^{(el)}(t).
   \label{eq:fulldyn}
\end{equation}
Here, $J_i$ and $D_i$ denote the inertia and damping constant, respectively, $P_{i}^{(in)}$ is the effective real power injection at node $i$, and $P_{i}^{(el)}$ is the real power exchanged with the grid. Throughout the paper we assume that the power is balanced such that $\sum_i P_{i}^{(in)} = 0$. We neglect transmission losses as Ohmic resistance is typically small in high-voltage transmission grids. Hence, the real power exchanged with the grid can be written as
\begin{align}
   P_{i}^{(el)}(t) &= \sum^N_{j=1} P_{ij}(t)  
   = \sum^N_{j=1} K_{ij} \sin \left( \delta_i - \delta_j \right),
\end{align}
where $P_{ij}$ is the real power flow on line $(i,j)$. The `coupling strength' $K_{ij}=K_{ji}$ is determined by the susceptance of the respective transmission line (being zero if no line exists), and the voltage level of the grid \cite{witthaut2022collective}. In the context of network science, equation \eqref{eq:fulldyn} is commonly referred to as the second-order Kuramoto model or Kuramoto model with inertia \cite{acebron2005kuramoto}.

In the simulations we first find a steady state of the equations of motion \eqref{eq:fulldyn} for the pre-failure grid. This is used as the initial state for the following simulations. Then we select a transmission line $(i,j)$ and remove it from grid by setting $K_{ij}$ to zero. We simulate the dynamics using \emph{PowerDynamics.jl}~\cite{PowerDynamics2022} and assess stability in term of the final state: A line is classified as stable if all nodal frequencies $\dot \delta_i(t)$ relax back to zero, and critical otherwise. This procedure is repeated for all lines in the respective grid.

\subsection{Power grid datasets}

\begin{table}[tb]
\caption{Synthetic grid data sets and their properties: $K_{ij}$ is the line capacity, $|P_C|$ is the effective power demand of a consumer node. The generator power $|P_G|$ then follows from the generator-consumer ratio $N_G$:$N_C$ since power is balanced. $N_G+N_C$ is the total number of nodes. Data sets $\US^P$ and $\US^P_B$ contain grids with $|P_C|$ = 1.5, 1.6, 1.7, 1.8, 1.9, 2.}
\resizebox{\columnwidth}{!}{%
\begin{tabular}{|l|l|l|l|l|l|l|l|l|}
\hline
                         & \textbf{$\US$} & \textbf{$\US_\circ$} & \textbf{$\US^P$} & \textbf{$\US^P_B$} & \textbf{$\US_{het}$} & \textbf{$\GB_{het}$} & \textbf{$\GB$} & \textbf{$\GB_{pert}$} \\ \hline
\textbf{$ K_{ij}$}       & $L_{ij}^{-1}$  & $L_{ij}^{-1}$        & $L_{ij}^{-1}$    & 4                  & $L_{ij}^{-1}$        & $L_{ij}^{-1}$        & $L_{ij}^{-1}$  & $L_{ij}^{-1}$         \\ \hline
\textbf{$|P_C|$}         & 2              & 2                    & {[}1.5, 2{]}     & {[}1.5, 2{]}       & 0.75                 & 1                    & 1              & 1                     \\ \hline
\textbf{$N_{G}$:$N_{C}$} & 1:1            & 1:1                  & 1:1              & 1:1                & 1:4                  & 1:4                  & 1:1            & 1:1                   \\ \hline
\textbf{$N_{G}+N_{C}$} & 50          & 50                & 50            & 50              & 50                & 120                & 120          & 120                 \\ \hline
samples                  & 43825          & 26734                & 126508           & 65568              & 21936                & 118400               & 29600          & 73556                 \\ \hline
\end{tabular}%
}
\label{t:data sets}
\end{table}

Training and evaluating machine learning models requires a data set of sufficient size. Real power grids are usually operated well within their margin of stability, such that desychronization events are rare if potentially catastrophic. Hence, it is impossible to use real data or models of normal operation dispatch scenarios in our study. Instead we resort to synthetic models which are close to actual grid topologies. In order to create data sets containing sufficient numbers of critical cases, the grids are designed as to be heavily loaded, and are not necessarily \emph{statically} $N-1$ stable, i.e. static stability is not guaranteed for all N possible single line failures.

Eight different data sets were produced, an overview is given in Table~\ref{t:data sets}. All test cases in a data set either have the British transmission grid topology (marked $\GB$), or the topologies were generated synthetically using the random growth model developed in \cite{schultz2014random}. In the latter case, parameters were chosen that produce topologies mimicking properties of the US transmission grid (marked $\US$). Next, we specify how many nodes in the grid act as effective generators ($P_{i}^{(in)}>0$) or effective consumers ($P_{i}^{(in)}<0$).
Nodes are then randomly assigned to one of the two classes. The ratio of the number of generators and consumers $N_G$:$N_C$ as well as the value of $|P_{i}^{(in)}|$ for the consumer nodes are given in Table~\ref{t:data sets} for all data sets. The transmission lines were either all modelled with the same effective coupling strength $K_{ij}$, or $K_{ij}$ was set to be inversely proportional to the line length as derived from node positions. The latter case treats all transmission lines as conductors of the same electrical conductivity. Obviously, $K_{ij}=0$ if two nodes $i$ and $j$ are not connected. Two data sets additionally contained grids of varying line loadings by rescaling node powers as shown in Table~\ref{t:data sets}. The data set marked by $\US_\circ$ had all dead ends removed by connecting them to the next closest node. Finally, the data set marked by $\GB_{pert}$ was created by adding randomly drawn perturbations of sum zero to every node power $P_{i}^{(in)}$ of a reference grid. In some case no steady state could be found for the pre-failure grid. These cases were discarded. The total numbers of simulated line failures vary by data set, see Table~\ref{t:data sets}.

\subsection{Machine learning models}
\label{sec:models}

For our predictions we use tree-based machine learning models because they provide state-of-the art performance for many applications \cite{ke2017lightgbm} as well as a high level of explainability \cite{lundberg2020local}. The output of each model is a number $y\in[0,1]$, which can be interpreted as the probability of a line failure leading to a desynchronisation. This number can be used to derive a classification by a simple thresholding procedure. 

We compare the performance of state-of-the-art gradient boosted tree (GBT) models~\cite{ke2017lightgbm} to gradient boosted stump model (stumps) and a simple decision tree (DT) model to find out how much complexity is needed to maximise performance. Here, a stump is a tree made up of only a root node and two leaf nodes. All models were subjected to hyper parameter optimisation via random search.

Tree-based models can be made transparent by different XAI methods. We use SHapley Additive exPlanations (SHAP)~\cite{lundberg2020local}, in particular for the quantification of feature importance. The SHAP value quantifies how strongly a feature influences a given prediction made by a model. Averaging over all predictions then yields the global feature importance, i.e. how much a given feature contributes to a model overall. To reduce model complexity and increase interpretability we try to find an ``optimal" model of maximal performance and minimal dimensionality. Since evaluating all possible feature combinations is computationally prohibitive, we perform recursive feature elimination (RFE) in which the feature with the smallest feature importance is removed recursively from the model. Note that RFE does not necessarily lead to the optimal feature combination due to limitations in importance attribution~\cite{ShapleySelection}.

For each data set $80\%$ of the samples are used for training and the remaining $20\%$ for testing. We use 4-fold cross validation, i.e. the performance is calculated by averaging over four different train-test splits.
Performance was assessed across multiple metrics, which generally agreed. For conciseness, in this paper we therefore limit ourselves to the Average Precision (AP) as a numerical and the detection error trade-off curve (DET) plot \cite{martin1997det} as a graphical performance metric. The AP is a well-suited metric for our purposes, as it quantifies the ability of an algorithm to rank samples by relevance, i.e. desynchronisation risk in our case. In potential applications cases machine learning would be combined with detailed numerical simulations. The output of the machine learning model would then be used to narrow down the vast amount of possible contingencies to those that should be investigated further via simulations.

\section{Metrics of link importance and redundancy}
\label{sec:metrics}

A key idea of this study is to combine machine learning with concepts from graph theory and network science which are used to engineer features that are interpretable and have a high predictive power, as for instance measures of redundancy or centrality.
Features are computed from the grid's topology (adjacency matrix), the electric properties ($P_i^{\rm in}$ and $K_{ij}$) and the pre-outage state (phase angles $\delta_i$ and flows $P_{ij}$).
In the following we introduce the engineered features; a summary is given in table \ref{table:feature_list}. We note that some of the features are widely used in network science, hence we introduce them very briefly. Furthermore, we will provide a first assessment of their predictive power using uni-variate models. 

\subsection{Definition of the features}

\begin{table}[tb]
\caption{List of the engineered features used to predict desynchronisation. Features for which no symbol is listed showed bad performance and are omitted in Figure~\ref{fig:singe_feat_perf} for conciseness. All features were used as features in the machine learning models.
}
\centering\setcellgapes{0.5pt}\makegapedcells \renewcommand\theadfont{\normalsize\bfseries}%
\begin{tabular}{|l|l|}
\hline
\textbf{Feature name/definition}       & \textbf{Symbol}              \\ \hline
redundant capacity ratio \cite{witthaut2016critical}       & $r_{ab}$              \\ \hline
widest path redundant capacity ratio    & $r^{w}_{ab}$          \\ \hline
shortest path redundant capacity ratio   & $r^{\sigma}_{ab}$     \\ \hline
redundant capacity \cite{witthaut2016critical}              & $K^{red}_{ab}$        \\ \hline
widest path redundant capacity           & $K^{red,w}_{ab}$      \\ \hline
shortest path redundant capacity          & $K^{red,\sigma}_{ab}$ \\ \hline
response theory pred. max. load \cite{witthaut2016critical}           & $\lre$              \\ \hline
LODF pred. max. load \cite{guler2007generalized}             & $\ldc$              \\ \hline
$\sqrt{\rc^2+(\ldc)^2}$ \cite{witthaut2016critical} & $\rl$                        \\ \hline
max. load at operation point                 & $l_{OP}$              \\ \hline
load on failing line                & $l_{ab}$              \\ \hline
flow on failing line                & $|P_{ab}|$            \\ \hline
phase cohesion \cite{dorfler_synchronization_2013}      & not shown    \\ \hline
post failure phase cohesion \cite{dorfler_synchronization_2013} & $\cohesPost$           \\ \hline
$\max_i|P_i| - \sum_j K_{ij}$ &         $b_{ab}$         \\ \hline
$\max_i|P_i| / \sum_j K_{ij}$ &       not shown           \\ \hline
rerouting resistance distance     & $X^{re}_{ab}$         \\ \hline
weighted rerouting distance           & $X^{\sigma}_{ab}$     \\ \hline
normalised edge current betweeness \cite{brandes2005centrality}  & $\epsilon^{CB}_{ab}$  \\ \hline
pre failure algebraic connectivity          &        not shown    \\ \hline
post failure algebraic connectivity        &    not shown \\ \hline
algebraic connectivity loss       & $\Delta\lambda_{2,ab}$     \\ \hline
rerouting distance              & $d^{re}_{ab}$         \\ \hline
edge betweenness             & $\epsilon_{ab}$       \\ \hline
edge connectivity             & $\tau_{ab}$           \\ \hline
edge k core               &     not shown       \\ \hline
\end{tabular}
\label{table:feature_list}
\end{table}

The most simple features describing a failing line $(a,b)$ are the flow $|P_{ab}|$ and the load $l_{ab} =  |P_{ab}|/K_{ab}$ in the pre-outage state. Intuitively, we expect that the failure of a strongly loaded line will have a stronger impact than a weakly loaded line.

A variety of measures of connectivity, redundancy and centrality were introduced in the context of network science. The \emph{edge connectivity} $\tau_{ab}$ of a link $(a,b)$ is defined as the number of edge independent paths between the nodes $a$ and $b$ and provides an elementary measure of redundancy \cite{Newm10}. If $\tau_{ab}=1$, then the removal of the edge $(a,b)$ will disconnect the grid and almost surely lead to a desynchronization. 
The \emph{algebraic connectivity} or Fielder value $\lambda_2$ is of particular interest in flow networks~\cite{Fied73,dorfler2012exploring}. It is defined as the smallest non-zero eigenvalue $\lambda_2$ of the graph Laplacian matrix $\mathbf{Y} \in \mathbb{R}^{N \times N}$ \cite{Fied73}
\begin{equation}
    Y_{ij}   =\left\{\begin{array}{l l }
      -K_{ij} & \; \mbox{if $i$ is connected to $j$},  \\
      \sum_{\ell} K_{i\ell} & \; \mbox{if $i=j$},  \\
      0     & \; \mbox{otherwise}.
  \end{array} \right. \label{eq:Laplacian}
\end{equation}
We evaluate this eigenvalue before and after the removal of the line $(a,b)$ (pre- and post-failure grid) and compute the difference $\Delta\lambda_{2,ab}$ as a measure for the loss of connectivity.

The \emph{edge betweenness} measures the centrality of an edge \cite{Newm10}. The original version $\epsilon_{ab}$ is defined by computing the shortest path between all pairs of nodes in the network and counting how many of these paths cross the edge $(a,b)$. Here, we also use the flow-based version $\epsilon^{CB}_{ab}$ where shortest paths are replaced by current flows \cite{Newman2005}. The \emph{coreness} of a line has been shown to be related to its vulnerability~\cite{yang2017small}. The $k$-core of a graph is its maximal subgraph that contains only nodes of degree $k$ or more. The coreness of a node $a$ is defined as the highest $k$ for which $a$ is part of the $k$-core and the coreness of a line is the smaller of the corenesses of its terminal nodes.

If an edge $(a,b)$ fails, its flow must be rerouted via alternative paths. The properties of theses paths, especially their length, may thus be relevant to determine the impact of the failure. The \emph{rerouting distance}  was introduced in \cite{strake2019non} to predict flow changes in linear flow networks resulting from line failures. The rerouting distance of two lines is defined as the length of the shortest loop that contains both lines. Here we introduce the rerouting distance of a single line $(a,b)$ as the shortest loop that contains the line. Up to an offset of one, this is equivalent to the to the geodesic distance between $a$ and $b$ in the post failure grid.
Here, we extend this definition to incorporate line properties, interpreting $K_{ij}^{-1}$ as the effective resistance of an edge $(i,j)$. We then define the \emph{weighted rerouting distance} $X_{ab}^{\sigma}$, as the effective series impedance of the weighted shortest path in the post failure grid.
We further define the \emph{rerouting resistance distance} $X^{re}_{ab}$ of a line $(a,b)$ as the resistance distance \cite{klein1993resistance} between $a$ and $b$ in the post failure grid.

\emph{Line outage distribution factors} (LODFs) are widely used in power system stability analysis~\cite{wood2013power, kaiser2021network, Zocca}. Assuming small phase angle differences between connected nodes the equations describing the steady state are linearized, $\sin(\delta_i-\delta_j) \approx \delta_i-\delta_j$, which allows to compute the post-failure steady state analytically, implicitly assuming that such a steady state exists~\cite{Guo09,strake2019non}.
To the same end, in \cite{manik2017network} \emph{linear response theory} was applied to the second-order Kuramoto model by linearizing the sine around the pre-failure steady state.
In our context the maximal predicted line loading $\max_{ij} |P_{ij}|/K_{ij}$ in the post-failure grid is of particular importance. A value greater than one hints at an overload and thus a loss of stability. This quantity is denoted as $\ldc$ (computed using LODFs) and $\lre$ (linear response theory), respectively. 

The maximal line loading at the pre failure operation point, $\max_{(i,j)\in E} \sin \left( \delta_i - \delta_j \right)$ can be used as a proxy for the overall line loading. We denote it as $l_{OP}$.

A simple necessary condition for the existence of a steady state of equation \ref{eq:fulldyn} is given by $|P_i^{(in)}| \le \sum_j K_{ij}$. Hence we consider the following quantities, evaluated after the removal of edge $(a,b)$, as potential features  
\begin{equation}
    \max_i |P_i^{(in)}| - \sum_j K_{ij}, \qquad
    \max_i |P_i^{(in)}| / \sum_j K_{ij}
\end{equation}
Dörfler et al~\cite{dorfler_synchronization_2013} introduced a synchronisation condition for the existence of a stable steady state with maximal phase difference $\gamma$ between any two connected oscillators. Adapted to our problem the criteria reads:
\begin{equation}
    ||\mathbf{Y}^{\dagger}\vec{P}||_{\epsilon,\infty}\leq \sin(\gamma),
\end{equation}
where $\mathbf{Y}^{\dagger}$ is the pseudoinverse of the graph Laplacian matrix, $\vec P=(P_1^{(in)},\ldots,P_N^{(in)})^\top$, and
$||x||_{\epsilon,\infty}=\max_{i,j\in E}|x_i-x_j|$. In the following we use the left hand side of the equation as an input feature and refer to it as the \emph{cohesiveness} $\rho$. The respective value in the post-failure grid, i.e. after removing the edge $(a,b)$ from the Laplacian, is denoted as $\cohesPost$.

\begin{figure*}[tb]
\centering
  \begin{minipage}[c]{0.63\textwidth}
    \includegraphics[width=\textwidth]{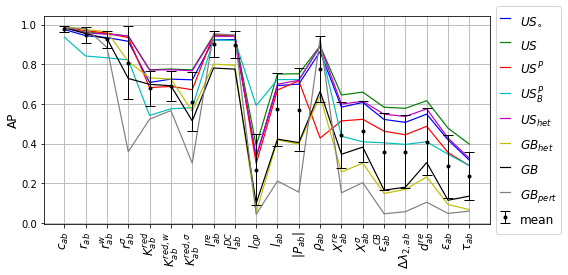}
  \end{minipage}\hfill
    \begin{minipage}[c]{0.36\textwidth}
    \caption{
    Performance of uni-variate threshold models for different engineered features.
    The performance is quantified by the average precision (AP) score, evaluated for the eight data sets summarised in table \ref{t:data sets}. Black bars indicate the mean and standard deviation over the data sets. Note that the performance might differ across different data sets not only because of a features strengths and weaknesses but also because a data set can be intrinsically easier or harder to predict. 
    Features take more information into account from right to left: first purely topological features, then including electrical properties, finally also the pre-failure steady state. 
    We find that features derived from the redundant capacity and the predicted maximum line load achieve the highest scores. They are also the most consistent over all data sets. Purely topological features show bad performance.}  
    \label{fig:singe_feat_perf}
    \end{minipage}
\end{figure*}

The \emph{redundant capacity} $K^{red}_{ab}$ of a line $(a,b)$ quantifies its redundancy in the pre-failure steady state of the grid \cite{witthaut2016critical}. This measure is inspired by graph theoretical flow problems, in particular the min-cut max-flow theorem \cite{Ahuj13}. It is defined as the total additional flow that the grid can transport from $a$ to $b$ by using all possible paths that don't include line $(a,b)$ itself.
Since $K^{red}_{ab}$ will be used to assess the impact of a line failure, we must take into account that lines typically carry a flow before the failure. Therefore, we do not ask how much flow a line $(i,j)$ can carry in total (given by the coupling constant $K_{ij}$)) but how much it can carry \emph{in addition to the pre-failure state}. Accordingly, we introduce a residual network $G^{res}$ in which lines are described by the residual capacity $K^{res}_{ij}=K_{ij}-P_{ij}$. Since $K^{res}_{ij} \neq K^{res}_{ji}$ we must view $G^{res}$ as a directed graph, where every line is represented by two directed edges with different capacity. The redundant capacity of the line $(a,b)$ is then obtained by removing the line $(a,b)$ from $G^{res}$ and then computing the maximum $(a,b)$-flow via the Edmonds-Karp algorithm \cite{Ahuj13}. By the virtue of the min-cut max-flow theorem, the redundant capacity $K^{red}_{ab}$ equals the minimum capacity of all $(a-b)$-cuts in the $G^{res}$ and thus identifies the bottlenecks between the nodes $a$ and $b$. 
Furthermore, we relate the redundant capacity to the amount of flow that has to be rerouted, which is given by the pre-outage flow on the failing line $|P^{\rm pre}_{ab}|$. The \emph{redundant capacity ratio} $\rc = |P^{\rm pre}_{ab}|/K^{red}_{ab}$ has a high predictive power for the impact of line failures as previously shown in~\cite{witthaut2016critical}.
Furthermore we consider a combination of the redundant capacity ratio and the maximum load predicted from LODFs, $\rl=\sqrt{\rc^2+(\ldc)^2}$ 
\cite{witthaut2016critical}.

The graph theoretical max flow provides an upper bound for real power flows, but not the actual value. Hence we introduce two variants of the redundant capacity $K^{red,\sigma}_{ab}$ and $K^{red,w}_{ab}$, which do not take into account all possible paths but only the shortest path ($\sigma$) path or the widest path ($w$) from $a$ to $b$, respectively. Here, the widest path is defined as the single path with the widest bottleneck, i.e. the largest value of $\min_{ij \in {\rm path}}K^{res}_{ij}$. As before, we additionally define the corresponding ratios $r^{\sigma}_{ab}$ and $r^{w}_{ab}$.

\subsection{Initial evaluation of the features}

We now provide a first assessment of the predictive power of the engineered features in terms of simple uni-variate prediction models. For each feature, a model is set up as follows. Given a threshold $h$ an edge is predicted as critical if the metric exceeds $h$ and predicted stable otherwise. By varying the value of $h$ one can derive the precision recall curve and thus the AP.
Figure \ref{fig:singe_feat_perf} shows the AP scores for the different features and all data sets. Note that for conciseness some of the features  introduced before that performed badly are not shown.
The features are ordered according to what information they take into account: from full information (topology, electric properties, and pre-failure state), over topological and  electrical properties, to purely topological.

The features showing the best performance are $\rl$, $r_{ab}$, $\rw$, $\rs$, $l_{re}$ and $l_{DC}$ , with $\rl$ clearly scoring the highest. Besides being strong predictors, they are also the most consistent, showing a comparably low performance variance. The full redundant capacity ratio $r_{ab}$ outperforms the related $\rw$, $\rs$. The score of $\lre$ and $\ldc$ is almost identical on all metrics and data sets. All of the aforementioned features quantify the impact of the line failure on the flows.
Notably, the next best feature, $\cohesPost$ also falls into this category as it also provides an estimate for the maximum line loading in post-failure grid. 

The ten best performing features all take into account the full state of the grid. The next best features either take into account electrical and topological information or the pre-failure steady state. The best performing purely topological features is $\dist$, which is outperformed by no less than sixteen other features. Still, it outperforms more common and complex topological features including the current flow centrality. This is likely due to the fact that rerouting features are more specific to the problem than centrality measures.

As described above, categorical predictions are derived by setting a threshold $h$. We find that the optimal threshold is mostly constant between different data sets for the best performing features. This also hints at a potential for generalisation as no further knowledge about a power grid is needed to gauge whether a failure might be critical or not. This is not given for other features. A very low impedance distance for instance does not prevent desynchronisations if the grid is already loaded to the point of failure.

We conclude that the best performing features are all based on network flows -- either employing a graph theoretical perspective or linearizing the steady state equations.

\section{Machine learning critical lines}
\label{sec:ml-results}

\subsection{Predictability of network desynchronization}

\begin{figure}[tb]
    \includegraphics[width=\linewidth]{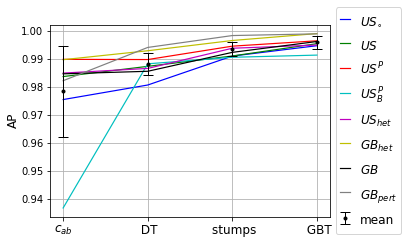}
    \caption{
    Performance of machine learning models averaged of four train-test splits.
    We show the average precision (AP) score of the decision tree (DT), gradient boosted stumps (stumps) and gradient boosted trees (GBT) models with $\rl$ as benchmark for all data sets. Black bars indicate the mean and standard deviation over the data sets. The GBT model consistently outperforms the other models.
    }
    \label{fig:AP_optimal}
\end{figure}

We now proceed to the main goal of our study, the combination of modern machine learning models and network science methods. The features introduced in the preceding section are used as inputs in models using gradient boosted trees (GBT), gradient boosted stumps (stumps) and a simple decision tree (DT). The feature $\rl$ is excluded here because it is already a combination of the two features $\rc$ and $\ldc$. Instead, we use a univariate model based on $\rl$ as a benchmark.

Figure~\ref{fig:AP_optimal} shows the performance of the optimal models for all data sets. We find that all machine learning models outperform the univariate benchmark $\rl$, that is, combining features improves the predictability. The performance increases with model complexity, with GBT models reaching an average precision of more than $0.996$ averaged over all data sets.
We conclude that the proposed approach of integrating capable network features via machine learning is highly effective to assess the network robustness.

\subsection{Reducing model complexity and identifying key features}

\begin{figure*}[tb]
  \begin{minipage}[c]{0.65\textwidth}
    \includegraphics[height=5cm]{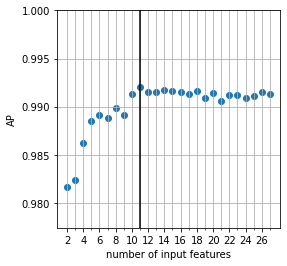}
    \includegraphics[height=5cm]{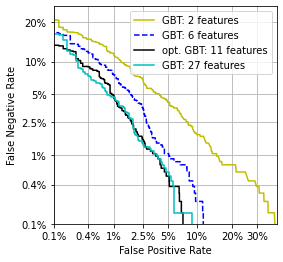}
  \end{minipage}\hfill
  \begin{minipage}[c]{0.35\textwidth}
\caption{Recursive feature elimination decreases model complexity and improves explainability. 
Left panel: Average precision score of the GBT model during recursive feature elimination. The vertical line marks the selected "optimal" model. Additional features do in theory not harm GBT performance, small differences occur by chance.
Right panel: DET curves for GBT models with different numbers of input features. The optimal and the 27 feature model perform almost identically confirming the choice of the optimal model. As evident from the 6 feature and 2 feature model eliminating more features leads to significant performance loss.
}
\end{minipage}
\label{fig:feat_elimination}
\end{figure*}

Tree-based machine learning models perform inherent feature selection which can be used to identify features with the highest predictive power. We apply recursive feature elimination to reduce model complexity and to improve the interpretability of the models. Figure~\ref{fig:feat_elimination} shows how the average performance over 4-fold cross validation evolves as features are gradually removed. In all cases, we find that many features can be discarded without loss of performance. Eliminating a feature does not harm performance if the feature is not a good predictor, or when the feature is redundant with respect to the remaining features. For stumps and especially decision trees we find the optimal number of features to be lower than for the GBT models. That is, the GBTs can better handle complex multi-dimensional inputs.

\begin{figure*}[tb]
    \centerline{\includegraphics[width=0.9\paperwidth]{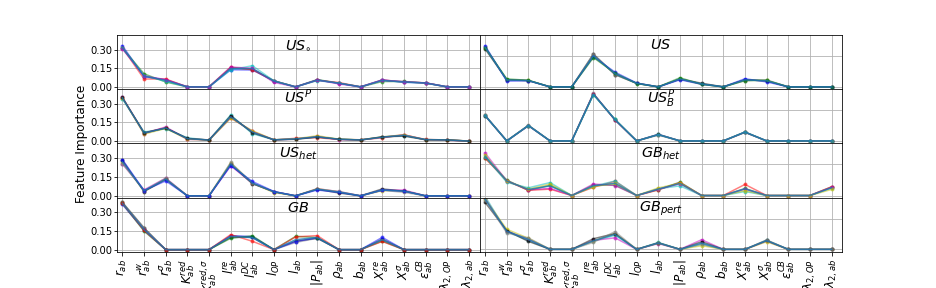}}
    \caption{
    GBT feature importance for all data sets. Results for 5 different train-test splits are shown, represented by different colouring. In most cases they strongly overlap, so differences are barely visible. Due to recursive feature elimination models have different input features and no model uses all features shown. Features not present in a model are represented by a zero feature importance. Redundant capacity ratio and $\ldc$ and $\lre$ dominate. Importances mostly fit the single feature performance shown in Fig.~\ref{fig:singe_feat_perf}.
    Correlated features compete for importance, so their importances turn out lower than expected from Fig.~\ref{fig:singe_feat_perf}. Notably $\cohesPost$ has an overall lower importance than $|P_{ab}|$ even though it shows much better single feature performance. Purely topological features barely contribute at all.
    }
    \label{fig:feat_imp}
\end{figure*}

We now investigate the individual features and their role in the machine learning prediction. Figure~\ref{fig:feat_imp} shows the feature importances, i.e. how much each feature contributes to the predictions on average, for the optimal GBT models. Feature importances are normalised so that they sum to one. Results are mostly consistent between data sets with data sets with higher similarity also showing more similar feature importance. We find that in general, feature importances are closely linked to the single feature performance shown in Fig~\ref{fig:singe_feat_perf}. That is, a feature with a high univariate predictive power will typically also show a high feature importance. The redundant capacity ratio and predicted line loading features have the highest feature importances. Purely topological features add very little value and are thus barely used by the models.

We note that two groups of features are based on similar concepts and are thus internally highly correlated. The two features $\ldc$ and $\lre$ are both based on a linearization of the power flows equations and thus partly redundant. Hence, models will use them interchangeably to a certain extent.
Similarly features based on the full redundant capacity and its shortest and widest path variants are to some some degree redundant and interchangeable.

Another outlier in the correlation between single feature performances and feature importance is $\cohesPost$. Despite being a good univariate predictor, the GBT models barley rely on it. The post failure cohesiveness $\cohesPost$ is a proxy for the maximal line load in the post-failure steady state. With $\ldc$ and $\lre$ though, two features exist that provide a similar proxy with substantially better performance. Hence, models will rather rely on $\ldc$ and $\lre$ than on  $\cohesPost$. 

\subsection{Generalisability}

Until now all models were trained and tested on different subsets of the same data, i.e.~the unknown test samples were drawn from the same underlying distribution as the training set. We now go a step further and analyse whether the learning translates between data sets. The rational of this is the following: A machine learning model will only be deployed in a real world application if it is expected to show good performance on real world data, especially in a high stakes situation. However, critical contingency events are rare so there is not enough real world data to test, let alone \emph{train} a machine learning model on.  Therefore, in an application case, training data will, by necessity, mostly be made up of synthetic test cases. We now use learning transfer between two different synthetic data sets as a proxy for learning transfer between synthetic and real word cases. Furthermore, a successful learning transfer would indicate that the machine learning model is not specific to one data set or network topology. This would substantiate the hypothesis that the model and the most important features describe intrinsic physical aspects of network stability and not just statistical correlations.

\begin{table}[tb]
\centering\setcellgapes{0.5pt}\makegapedcells \renewcommand\theadfont{\normalsize\bfseries}%
\caption{To assess the generalisability, models were trained and tested on different data sets. Every data set was used as a test set once.}
 \begin{adjustbox}{width=0.49\textwidth}
\begin{tabular}{|c|c|c|c|c|c|c|c|c|}
\hline
\textbf{train} & $\US$       & $\GB$ & $\US_B^P$ & $\US$     & $\GB_{het}$ & $\US_{het}$ & $\US$ & $\US_\circ$  \\ \hline
\textbf{test}  & $\US_\circ$ & $\US$ & $\US^P$   & $\US_B^P$ & $\US_{het}$ & $\GB_{het}$ & $\GB$ & $\GB_{pert}$ \\ \hline
\end{tabular}
\end{adjustbox}
\label{t:transfer}
\end{table}

\begin{figure}[tb]
    \includegraphics[width=\linewidth]{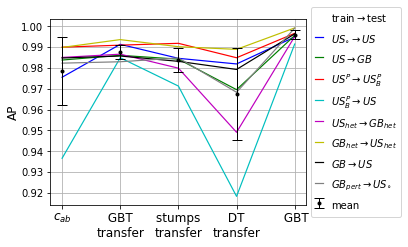}
    \caption{AP score of the transfer performance for the decision tree (DT), gradient boosted stumps (stumps) and gradient boosted trees (GBT) model with $\rl$ as benchmark. While not as good as the original GBT model the GBT transfer model clearly and consistently outperforms $\rl$.} 
    \label{fig:AP_transfer}
\end{figure}

To test the generalisability, data sets were paired so that for every data set a transfer model was chosen, see Table \ref{t:transfer}. Figure~\ref{fig:AP_transfer} shows the performance of the different transfer model types with $\rl$ and the original optimal non-transfer GBT model as benchmarks.
We find that learning transfer is possible in principle, albeit at a reduced performance. The performance of the GBT and the stumps transfer model lie between the two non-transfer benchmark models: higher than the uni-variate model $\rl$ and lower than the full GBT model. In contrast, the decision tree transfer models do not outperform the $\rl$ model.

The transfer performance does of course depend on the two data sets. In this study, the choices of respective training sets were arbitrary and not necessarily optimised to enable easy learning transfer. Hence, test and training data sets may have differing topologies, electrical properties, line loads and also different class frequencies. It thus stands to reason that a machine learning model trained on data from high-quality simulation models would offer much improved performance compared to uni-variate models based on single features such as $\rl$, $\lre$ or $\rc$.

\subsection{Interpreting the ML models}


\begin{figure*}[tb]
  \begin{minipage}[c]{0.65\textwidth}
    \includegraphics[height=4.5cm]{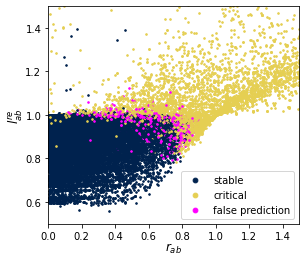}
    \includegraphics[height=4.5cm]{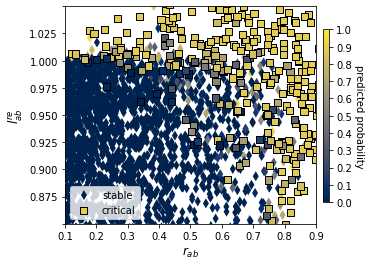}
  \end{minipage}\hfill
  \begin{minipage}[c]{0.35\textwidth}
    \caption{
    Relation between key input features and model outcomes for the $\US_\circ$ data set.
    Left: All samples of the data set projected into a two-dimensional feature space.
    Due to the high single feature performance of $\rc$ and $\lre$, stable and critical samples are mostly separated already in two dimensions. Most misclassified samples are found in the overlap region.
    Right: Magnification of the data around the overlap region. The colour code shows the probability output of the GBT model. In the overlapping region, the GBT model shows a significant improvement beyond the best prediction achievable by a two-dimensional model.
} \label{fig:prediction_analysis}
  \end{minipage}
\end{figure*}

We complete our study by a more detailed look on the relation of the key input features to the GBT model output. The left panel in Figure~\ref{fig:prediction_analysis} is a two-dimensional projection of the $\US_\circ$ data set, showing the two most important features $\rc$ and $\lre$. 
Those samples that were correctly classified by the machine learning model are coloured according to the outcome, while false predictions are coloured in red.
We find that while the two features already separate critical and stable samples with high accuracy, there is an are of overlap. By drawing on more features the optimal GBT model is able to reduce errors significantly beyond what would be possible using only those two features (see Fig.~\ref{fig:feat_elimination}). Nonetheless, most false predictions of the GBT model occur in this overlap region.

The right panel in Figure~\ref{fig:prediction_analysis} shows a magnification around the overlap region, with a colour mapping according to the probability output $y$ of the GBT model. As desired, critical samples generally have a higher probability output as stable samples. The figure elucidates the additional value of using additional features in the GBT model instead of relying on just the two best performing ones. Both classes overlap in the two-dimensional feature space spanned by $\rc$ and $\lre$. Still, the GBT model provides accurate predictions for many samples in this region. The additional features improve the models predictions, even though especially the best performing among them are very highly correlated to either $\rc$ or $\lre$.

\section{Conclusion and outlook}

Power grid system stability is an important and timely topic as large-scale outages can have catastrophic impacts. In this paper we improved upon known ways of predicting the dynamic stability of a power grid to line failures by combining classical power system tools, graph theory inspired feature engineering and interpretable machine learning models. Since the prediction methods applied are computationally cheap, they should be well suited to real world applications.

Using a second order Kuramoto model we simulated line failures for eight classes of synthetic grid models and classified whether they lead to a desynchronization event or not. We assessed the ability of 27 different features that quantify, among other, a lines load, connectivity, redundancy or centrality to correctly classify dynamic stability outcomes. We then used these features as input for different tree based machine learning models. After reducing model complexity by performing recursive feature elimination to improve interpretability we compared the performance of the different models. We found that especially gradient tree boosting models produce very good predictions reaching an average precision (AP) that exceeds $0.996$ when average over all data sets.

The engineered features contribute very differently to the models. In particular, we identified two classes of features that have a high predictive power and feature importance. First, a linearization in the spirit of line outage distribution factors allows to predict the maximum flow in the post-failure grid from the pre-failure state. Second, graph theory provides an upper limit for the ability of a grid to reroute power flows and thus a measure of redundancy for each line in the grid.

Features that are not specific to flow networks generally performed very badly and had very low feature importance. This is particularly true for purely topological features including connectivity, centrality or coreness. These findings are of particular relevance for studies of network robustness in network science which frequently refer to power grids as potential applications cases. Many classical studies focus on purely topological network properties, see, e.g.~\cite{albert2004structural,kinney2005modeling,wang2009cascade,buldyrev2010catastrophic}.
Based on our results, we conclude that purely topological metrics do not provide good predictors for the vulnerability of electric power grids, cf.~\cite{hines2010topological,korkali2017reducing}. 

Applications to real world grids require sufficient training data that can only be obtained from simulations. Hence, it is essential to generalise and transfer learned results from one system to another. To assess generalisability we tested the machine learning models on different data sets than they were trained on. The GBT model was still able to consistently outperform the best single feature predictor. We conclude that good performance should be achievable, especially since training data would be engineered to mimic potential critical real world contingencies, while our data sets were created to cover a wide range of grid properties.

In conclusion, we have demonstrated the potential of integrating machine learning models and network science metrics to assess the robustness of networked systems. 
A natural extension to our results would be to use larger and more realistic power system models. Our approach could also be applied to other stability risks, such as node failures or pertubations, and types, such as voltage stability, transient overloads or overload cascades. Training and deploying different models, transmission grid operators should be able to greatly improve their ability to identify potential risks in time, thus allowing early interventions where necessary and ultimately further improving power grid stability.

\section*{Acknowledgments}

We gratefully acknowledge support from the German Federal Ministry of Education and Research (BMBF grant no. 03EK3055B) and the Helmholtz Association via the Helmholtz School for Data Science in Life, Earth and Energy (HDS-LEE), Germany.

\bibliographystyle{IEEEtran}
\bibliography{references}

\begin{IEEEbiography}[{
\includegraphics[width=1in,height=1.25in,clip,keepaspectratio]{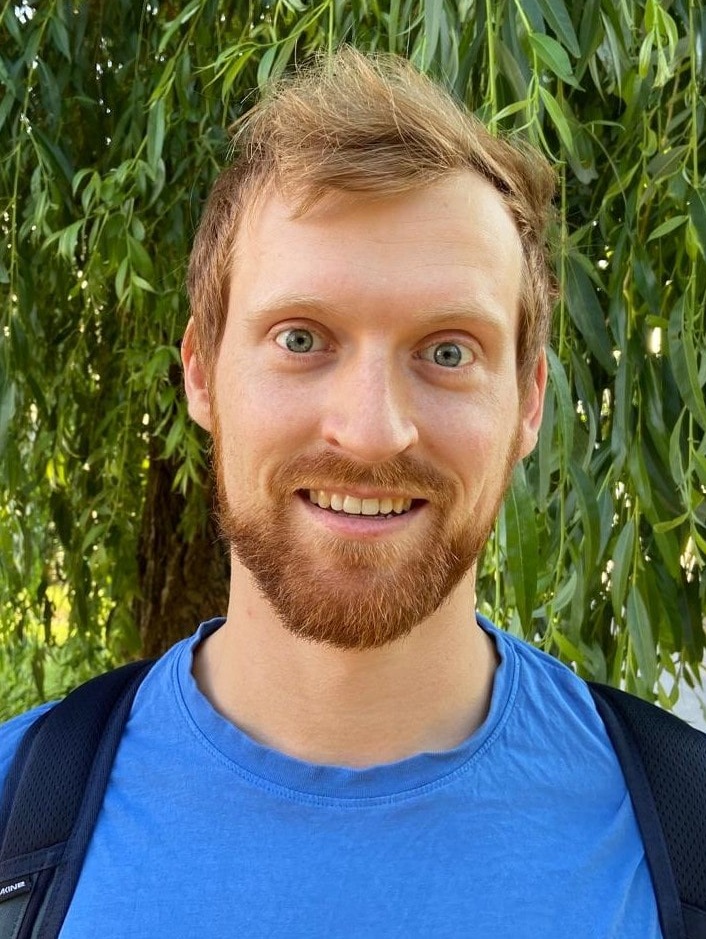}
}]{Maurizio Titz} obtained his B.Sc. and M.Sc. degrees in physics in 2019 and 2022. Currently, he is pursuing a Ph.D. degree in the University of Cologne and the Forschungszentrum Jülich, Germany.
\end{IEEEbiography}
\begin{IEEEbiography}[{
\includegraphics[width=1in,height=1.25in,clip,keepaspectratio]{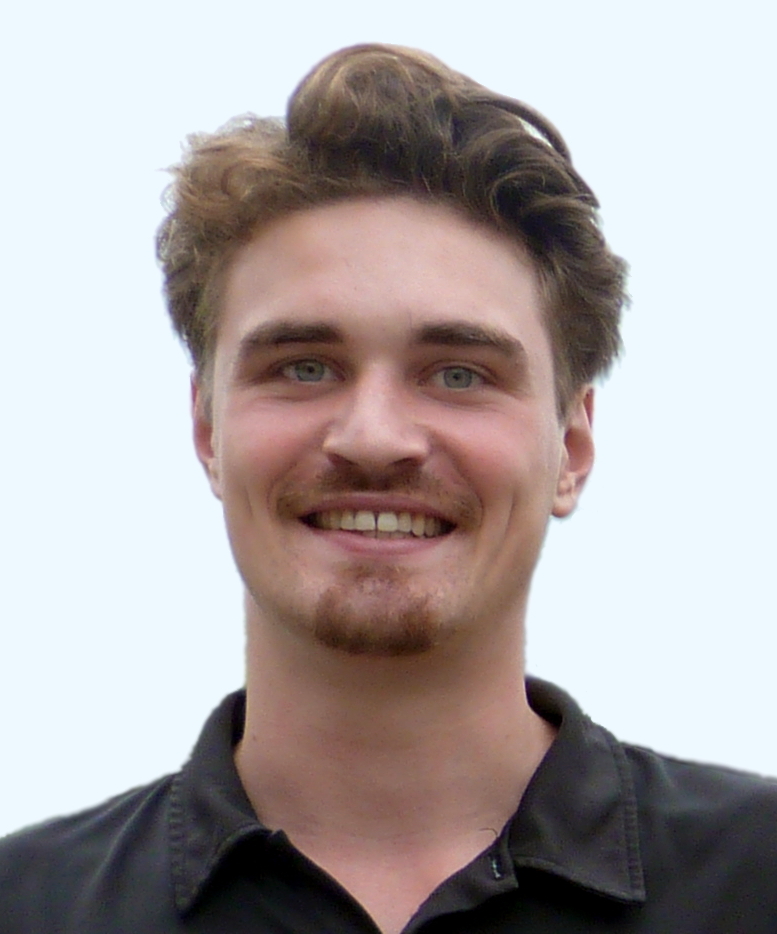}
}]{Franz Kaiser} obtained his B.Sc. and M.Sc. degrees in physics in 2015 and 2018, respectively, at the University of Göttingen and with the Max Planck Institute for Dynamics and Self-Organization in Göttingen, Germany. He received a Ph.D. degree in Physics from the University of Cologne, Germany, in 2021.
\end{IEEEbiography}
\begin{IEEEbiography}[{
\includegraphics[width=1in,height=1.25in,clip,keepaspectratio]{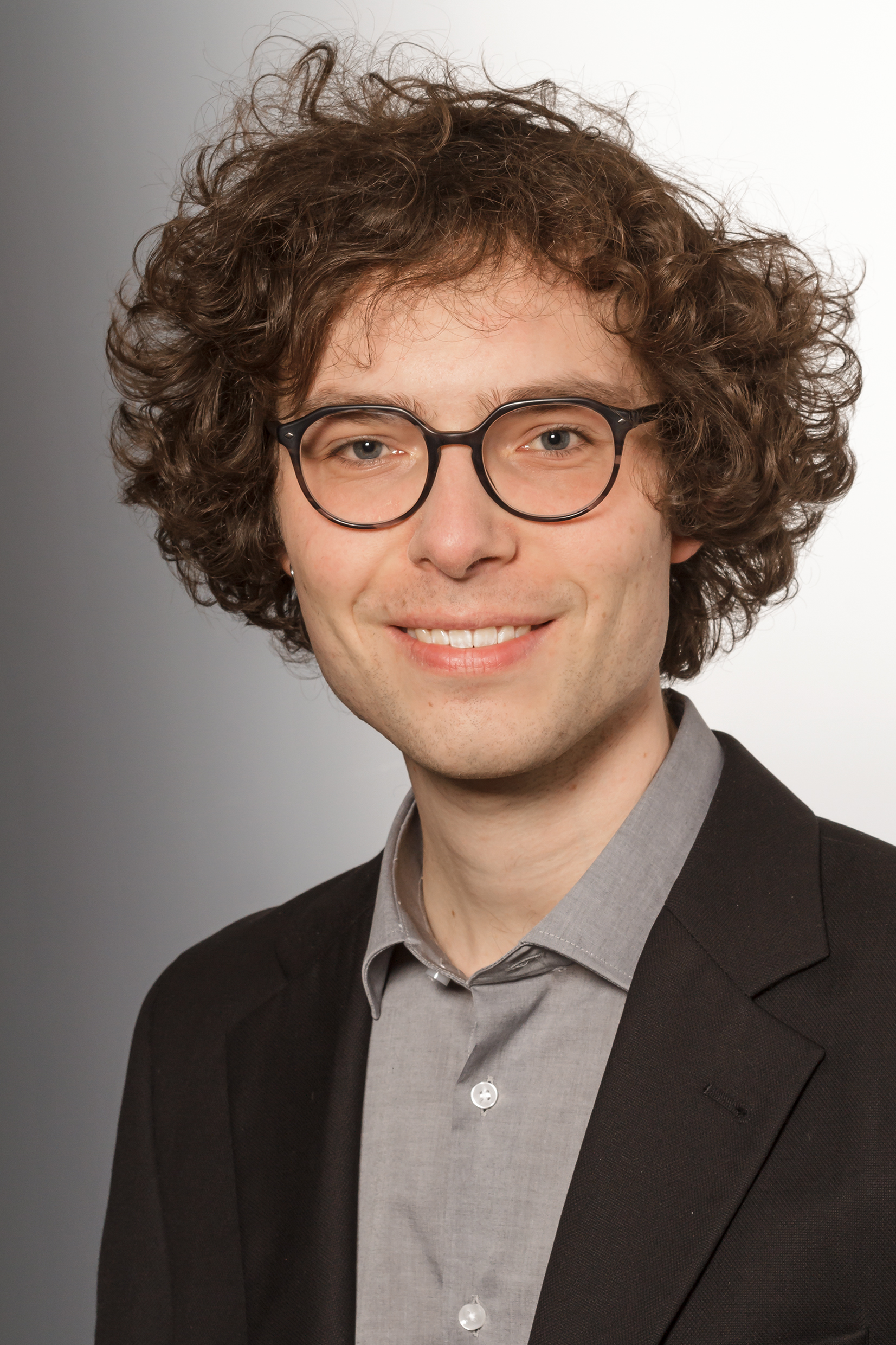}
}]{Johannes Kruse} received the B.Sc. and M.Sc. in physics at the Georg-August University of Göttingen, Germany in 2016 and 2019. During the final year of his studies, he wrote the Master's thesis at the Department of Engineering, Aarhus University, Denmark. Currently, he is pursuing a Ph.D. degree in the University of Cologne and the Forschungszentrum Jülich, Germany.
\end{IEEEbiography}
\begin{IEEEbiography}[{
\includegraphics[width=1in,height=1.25in,clip,keepaspectratio]{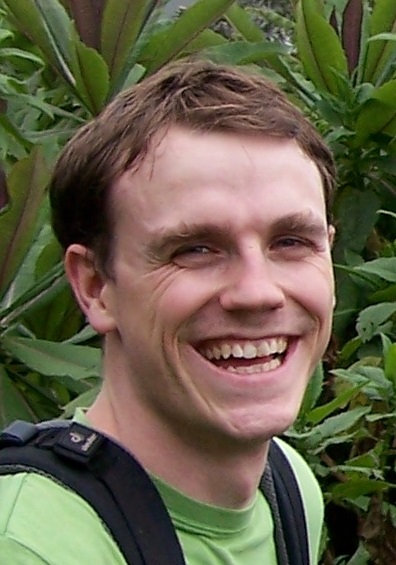}
}]{Dirk Witthaut} received his Diploma degree in Physics and his Ph.D. from the Technical University of Kaiserslautern, Kaiserslautern, Germany, in 2004 and 2007, respectively. He has worked as a Postdoctoral Researcher with the Niels Bohr Institute, Copenhagen, Denmark, and the Max Planck Institute for Dynamics and SelfOrganization, Göttingen, Germany. He has been a Guest Lecturer with the Kigali Institute for Science and Technology, Rwanda. Since 2014, he is leading a Research Group at the Forschungszentrum Jülich, Germany and he is a Professor with the University of Cologne. His research is focused on data and network science and applications to energy systems.
\end{IEEEbiography}

\end{document}